\documentclass[
 twocolumn,
 prl,
showpacs,
 amsmath,
 amssymb,
 aps,
]{revtex4-1}

\usepackage{graphicx}
\usepackage{epsfig}
\usepackage{dcolumn}
\usepackage{bm}

\begin{document}

\preprint{APS/123-QED}

\title{Structure and Response in the World Trade Network}

\author{Jiankui He and Michael W.\ Deem}

\affiliation{ \hbox{}Department of Physics \& Astronomy, Rice University,
Houston,Texas 77005, USA}

\date{\today}

\begin{abstract}
We examine how the structure of the world trade network has been shaped
by globalization and recessions over the last 40 years.
We show that by treating the world trade network as an evolving system,
theory predicts the trade network is more sensitive to evolutionary shocks and recovers
more slowly from them now than it did 40 years ago, due
to structural changes in the world trade network
induced by globalization.  We also
show that recession-induced change to the world trade network leads to
an \emph{increased} hierarchical structure of the global trade
network for a few years after the recession.
\end{abstract}

\pacs{89.65.Gh,89.75.Hc,89.75.-k}
\maketitle

Physical theory of evolution predicts that under certain
conditions, a changing environment  leads to
development of modular structure \cite{jun,he2009,Alon2005}.
The prediction depends only on 1) the dynamics of the response to change
being ``slow'' due to a glassy landscape, 2) presence of change, and
3) exchange of information between evolving agents.
Since the trade network is an
evolving system, this physics of evolution may be applied
to the world trade system, previously studied by network analysis \cite{hidalgo2007,barigozzi2010}.
We assume that condition 1 is satisfied for the world
trade network due to the complexities of inter-country relationships.
Condition 2 is satisfied by viewing
recessions as causing a change of the environment for
the dynamics of the world trade system.
Condition 3 is satisfied because
information flow naturally results from transfer of business
practices or material between countries.
 Thus, the theory of \cite{jun,he2009} allows us to make
three predictions: decreased modular
structure in the world trade network increases the sensitivity to recessionary
shocks, decreased modular structure decreases the rate of recovery, and recessions
themselves spontaneously increase modular structure of the world trade network.
All three predictions will be borne out by data.
These results are general predictions about how the detailed structural
parameters of the evolving economic system will organize.
Our theory shows that the modular and hierarchical structure formed in response to
environmental fluctuation increases the resistance to and
rate of recovery from perturbations.
The theory predicts that globalization, which reduces hierarchical
structure, should lead to increasingly large
recessions and decreased rate of recovery, in contrast to
standard economic understanding \cite{alesina2005}.

To apply the physical theory of evolution that describes
the spontaneous emergence of modularity in fluctuating environments
\cite{jun,he2009} to world trade,
we seek a mathematical representation
of hierarchy in the world trade network.
Identification of network motifs or modules is an active research field
in the physics of networks \cite{jun,newman2006,alon2007},
with the study of structure at multiple scales, i.e.\ hierarchy,  
somewhat more recent
 \cite{r26,clauset2008}.
In this paper, we treat the world trade data as defining a geometry in
trade space.  We project the trade topology onto the best tree-like
topology
representing the data.  The success of this projection in
representing the original geometry is used to define the
hierarchy of the original data.

We apply hierarchical clustering to construct the best
tree-like representation of the world trade
network.  Correlation between the distances
implied by the tree construction and the distances defined by the
original trade data is calculated.  This quantity is termed the
 cophenetic correlation coefficient (CCC) \cite{everitt2001}.
We will display the general trend of the CCC since 1969,
noting especially the increase of the CCC after each recession.
The magnitude of the CCC will be shown to
correlate with the ability of total world GDP to resist a recessionary shock.
Theory shows that this result is, in fact, causal, not simply a 
correlation, which is a major result here.


We focus on how global recessions, such as the 2008--2009 recession,
 have affected the structure of the world trade
network.   Modular structures arises in
the trade network, for example, because countries in a trade group
trade among themselves to a greater extent than with others. These
trade groups may interact with each other to form higher
level groupings.
The detailed reasons for an increase of hierarchy in the world
trade network are many:
perhaps protectionism for the domestic economy \cite{anderson2009},
or because long-distance trade seems costly during a recession.  
Standard arguments in economic theory suggest a decreased rate of recovery from recession for
trade networks with more modular structure \cite{alesina2005}.
We will see, however, that our theory
predicts that greater trade network structure increases both the resistance to
recessionary shocks and the rate of recovery from recessions.

\begin{figure}
\begin{center}
\epsfig{file=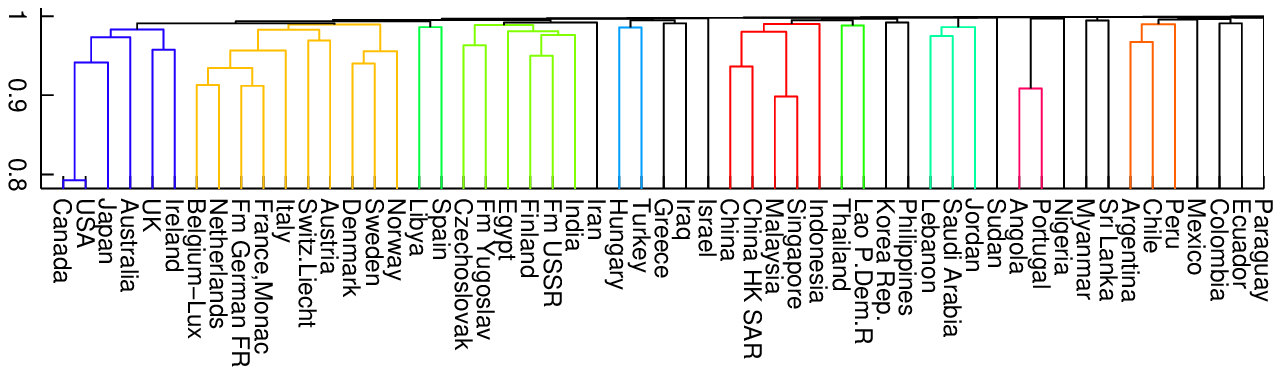,angle=270,width=3.in,clip=}
\epsfig{file=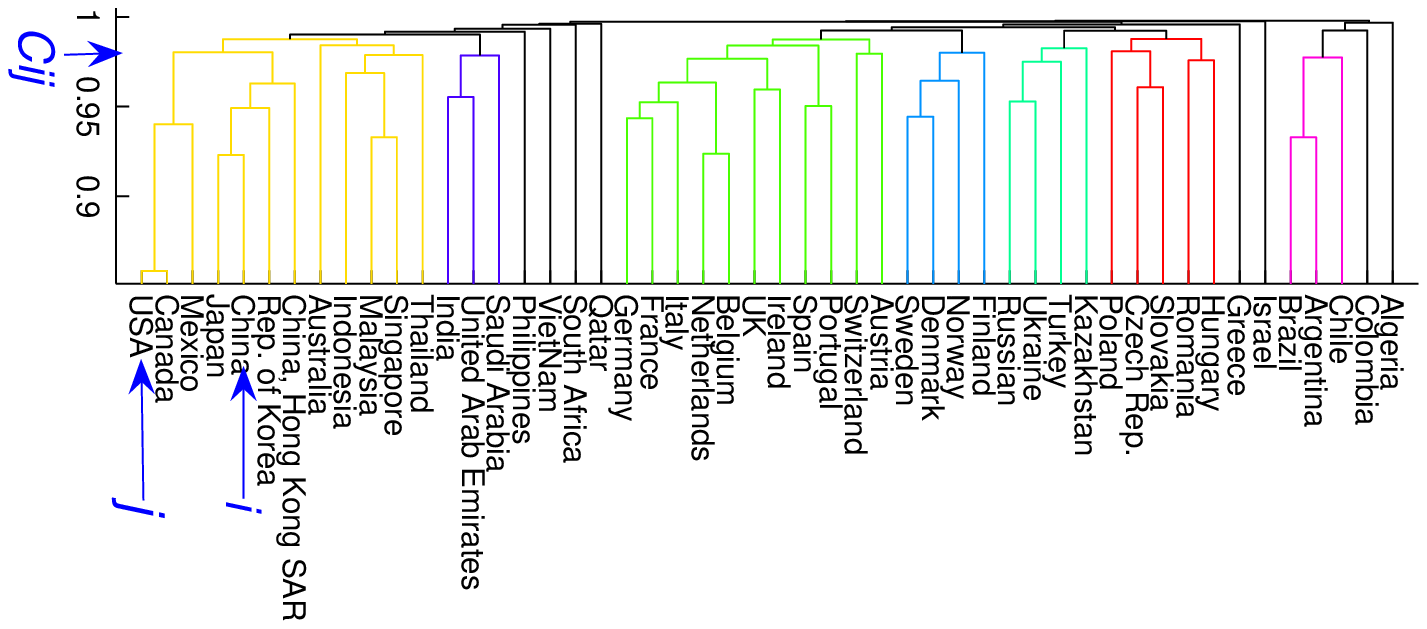,angle=270,width=3.in,clip=}\\
\caption{(color online) Dendrogram representation of trade
networks for selected countries at 1969 (top figure) and 2007 (bottom figure). Trade modules are marked by different colors. The structure of trade networks has changed greatly from 1969 to 2007. We plot only selected countries, because the figure becomes crowded if all
countries are plotted.
  \label{tree} }
\end{center}
\end{figure}

A hierarchical trade network occurs when countries with strong trade connections group into trade modules or regional trade clusters. A flat or non-hierarchical structure occurs when countries trade evenly with
 all other countries, and there are no regional trade modules in the trade network.
 We use the historical trade data from United Nation
database (Comtrade) from 1962 to 2007. We build the world trade
network with nodes representing countries and links representing the
trade value. We do not scale the trade volume by
the GDP, because small economic units should not
have the same weight as large economic units.
 First, a distance matrix is calculated from the
trade network matrix by $d_{ij} = M^* - M_{ij}$, where $M^* = \max(M_{ij})$.
Here, $M_{ij}$ is
trade value between two countries.
The average linkage hierarchical clustering
algorithm is applied to the distance matrix to produce the
tree-like dendrogram \cite{everitt2001}, see Fig.\ \ref{tree}. We define the tree-like structure  to have the most hierarchy.
Therefore, the amount of hierarchy can be measured by the likeness between
the original data and the best tree that is produced from original data by hierarchical clustering.
The CCC quantifies this likeness.
 The cophenetic matrix is generated
from the dendrogram. Its elements are the branch distance where two objects
become members of the same cluster in the dendrogram: for two nodes,
$ij$, we find the nearest common bifurcation point, and the branch length
for this point is the cophenetic element of these two nodes,
$c_{ij}$, see Fig.\ \ref{tree} for an example. The CCC is defined as
$CCC=[\sum_{i<j} (d_{ij}-\overline{d})(c_{ij}-\overline{c})]/\sqrt{[\sum_{i<j}(d_{ij}-\overline{d})^2][\sum_{i<j}(c_{ij}-\overline{c})^2]}$,
where $d_{ij}$ and $\overline{d}$ are the element and average of
elements of the distance matrix, and $c_{ij}$ and $\overline{c}$
are the elements and average of elements of cophenetic matrix,
respectively.
Hierarchical datasets have a high CCC value, and
nonhierarchical datasets have a low CCC value \cite{kaesler1972}.

\begin{figure}
\begin{center}
\includegraphics[scale=0.3]{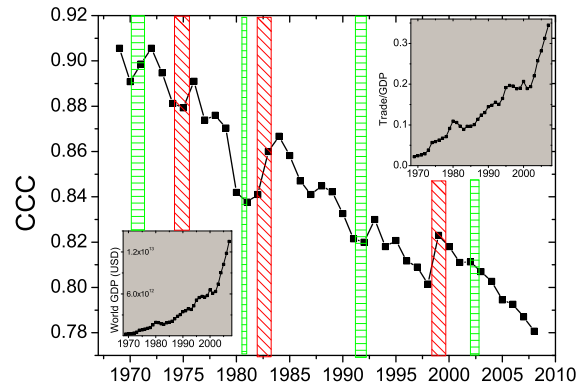}\\
 \caption{(color online) The CCC from 1969 to
2007. Data from 2008 are not yet available. 
Shaded rectangles marked the seven recessions.
Left and right borders
are positioned at the start and end of a recession, respectively,
according to US National Bureau of Economic Research.
During sharp recessions, the CCC increases
significantly, and during mild recessions, the CCC usually increases
mildly. The upper right insert is the ratio of total world trade to world GDP.
The lower left insert is the total world trade in units of US dollar.
  \label{ccc1}}
\end{center}
\end{figure}

A major factor affecting the world trade network over the last 40
years has been the process of globalization. Qualitatively, this
globalization has been expressed as a ``flattening'' of the world
\cite{Friedman}. Here, we use the CCC to measure how the
hierarchical structure of the world trade network has changed over
time. Large CCC values indicate higher hierarchy. 
The major trend of CCC with time in 
Fig.\ \ref{ccc1} 
is a reduction of
hierarchy as the ``flattening'' has taken place \cite{Friedman}. We
notice, however, that the CCC does not always decrease year by year.
We notice that
during and after each recession, marked on the figure,
the CCC value increases. The CCC values at the year after recession
are larger than that at the year before the recession ($p$-value $=0.003$ of Kolmogorov-Smirnov test for null hypothesis that they are from the same distribution with the same mean, and $p$-value $=0.0006$ for null hypothesis that CCC value before recession is larger than that after recession). This trend
is true both for the past 3 major recessions and for the past 4
minor recessions. The scale of increase of hierarchical structure
depends on the severity of recession. One possible reason for this
CCC trend during recessions is the increase of trade protectionism
during recessions. Also, regional integrations are greatly enhanced
during recessions, leading to increased regional imports \cite{Elliott2004},
which strengthens trade modules.
 Free trade promotes globalization and decreases
the hierarchy of the trade networks. But trade protectionism and regional integration, which is
common during recessions to protect domestic or regional economies by
restraining trade between countries, tends to reduce  trade between
countries in different trade modules.
Thus, recessions  may promote
the regionalization that enhances the modularity of the trade network.
One example is the Asian currency crisis of 1997, which lead to the
development of independent Asian monetary systems.

\begin{figure}
\begin{center}
\includegraphics[scale=0.3]{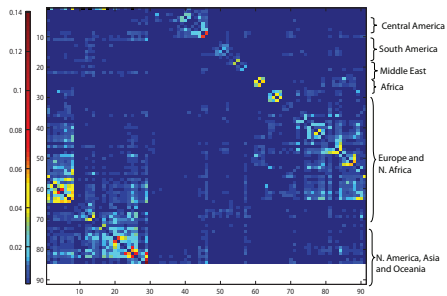}
\caption{(color online) The trade share matrix $S_{ij}=M_{ij}/\left(\sum_{m=1}^N
M_{im} +\sum_{n=1}^N M_{jn}\right)$ after hierarchical clustering between countries in 2007. We can see clearly several
modules: North American+Asian+Oceanic countries, European+ North African countries, Middle and South Africa,
Middle East, South America, and Central America.
  \label{tree2} }
\end{center}
\end{figure}

The CCC is a characterization of the world trade network that is
independent from the total amount of world trade.
 In the process of globalization, a country tends not only
to increase its total trade value, but also to trade with more
partners. The upper right insert of Fig.\ \ref{ccc1} shows the
typically increasing ratio of
world trade to GDP. Only the recessions of 1981, 1991, 1997
and 2001 lead to a decrease in the trade to GDP ratio, whereas the
CCC increased in all seven recessions. The increased hierarchical structure
appearing after all seven recessions in Fig.\ \ref{ccc1}, is
therefore, a sensitive correlate of recessions, and independent
of the trade to GDP ratio shown in the upper right insert of Fig.\ 
\ref{ccc1} and total trade volume shown in the lower left insert of Fig.\
 \ref{ccc1}. Measurement
of globalization by both hierarchical structure (CCC) and total trade provides
complementary information.

The CCC quantifies the development of hierarchical structure in the trade network
at multiple scales in an integrated way. The clustering of the world
trade network shows the modularity of global trade, see Fig.\
\ref{tree2}.
The development of regional trading partners occurs
simultaneously with globalization. By comparing the structure of trade network in 1969 
 and in 2007, we found that the increased trade among
Canada, United States, and Mexico as a result of NAFTA is one
example of a regional trading group.   Regional trade pacts among
the Middle East countries are other examples of regionalization.
 In general, free trade markets will develop modular structure at
multiple geographical scales.

\begin{figure*}
\epsfig{file=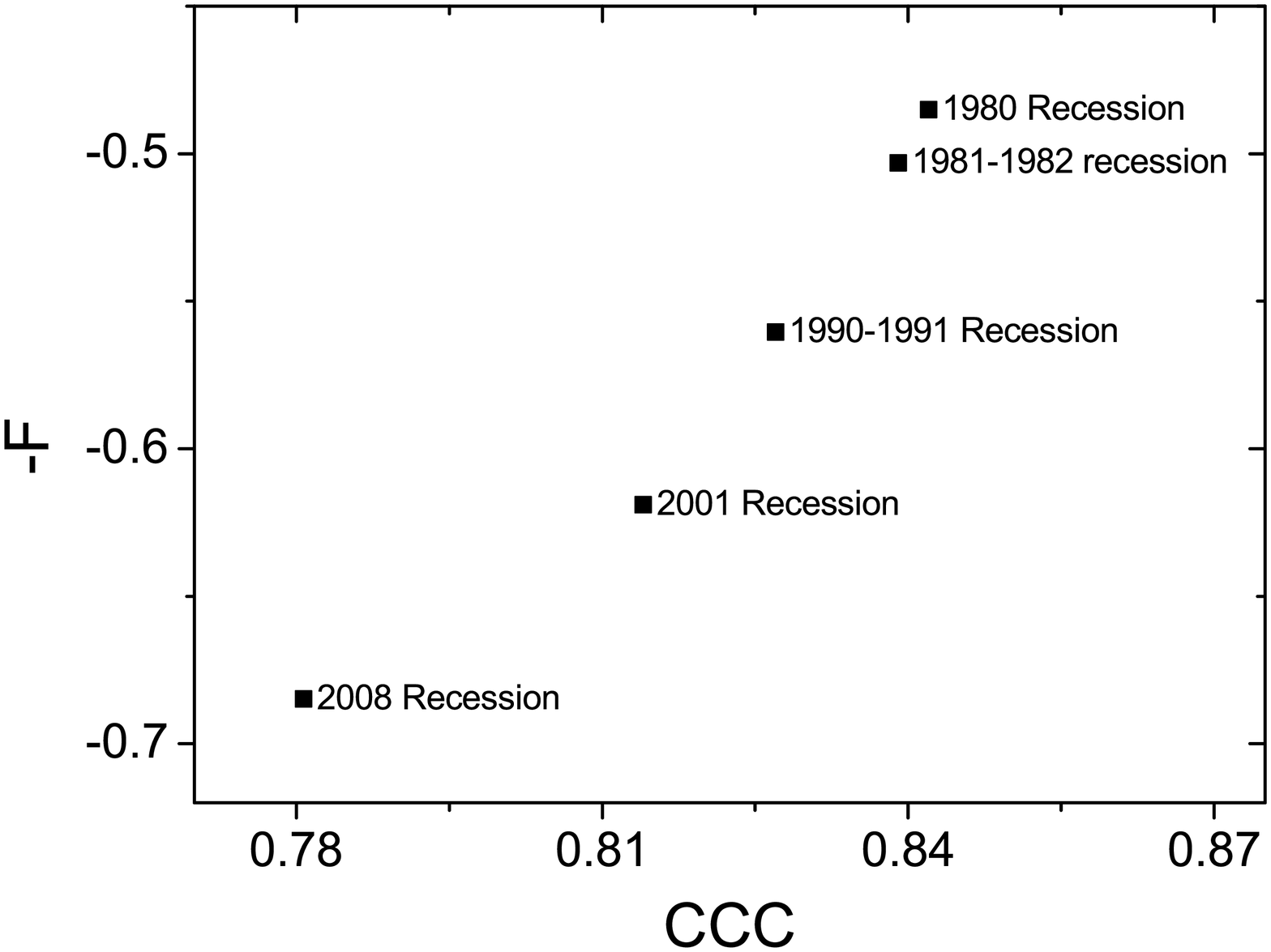,height=2in,clip=}
\epsfig{file=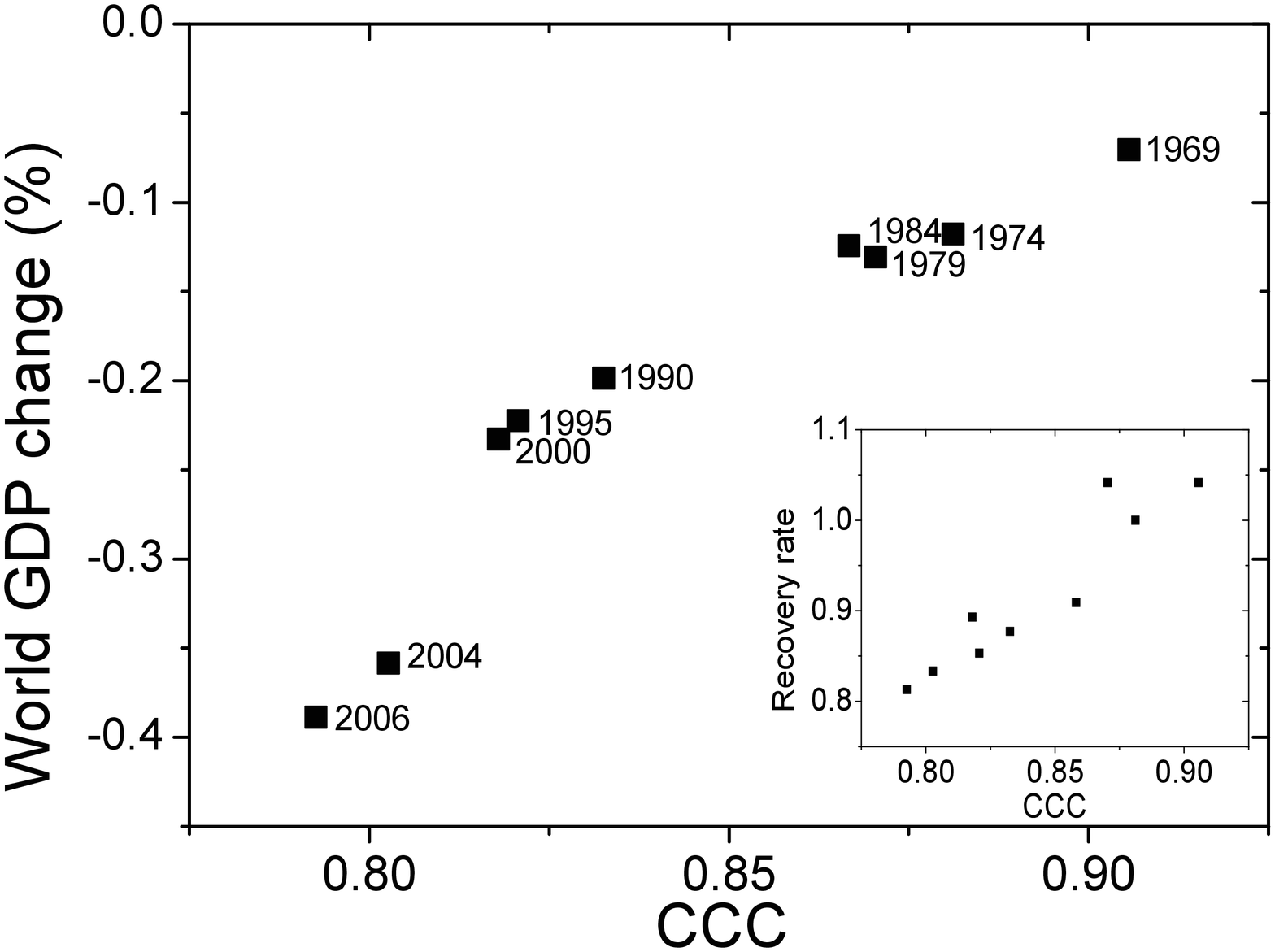,height=2in,clip=}\\
\hspace{0.1in}(a)\hspace{2.5in}(b)\\
 \caption{(a) The ratio of the total world excluding the USA GDP change (percentage) to the change of the USA GDP (percentage) in 5 recessions,
$F$ in the $y$ axis.
 The GDP change in one recession is defined as the
GDP decline from peak to trough. Quarterly GDP data are used
to find the peak of recession. The quarterly GDP data for 2001 and 2008 recessions are from \textit{OECD Stat.\ Extracts} (http://stats.oecd.org/). The quarterly GDP data for earlier recessions are estimated from annual GDP data.
  (b) Impulse response analysis of spread
of recession. In each year, initial values are set
to historical trade and GDP data.
A recession is assumed to
begin in the USA and spread to the rest of the world.
The world GDP change is plotted as a function of the CCC.
The reduction in the world GDP is greater when the CCC value is low.
Insert figure: The GDP recovery from recession can be well fit by
the relation $Y(t)\sim Y(\infty) - a\exp (-\lambda t)$.
Yearly recovery rates, $\lambda$, are shown versus the CCC.
In accord with theory, the recovery rate is positively correlated with the CCC.
  \label{sim} }
\end{figure*}

The ability of the trade system to respond to recessionary
perturbations is proportional to the hierarchical structure present,
i.e.\ increases with the CCC value, according to the evolutionary theory
of modular structure development \cite{jun,he2009}. That is,  the modular structure
that exists at multiple scales affects how recessions propagate in
the trade network, just as modular structure of person-to-person contacts
affects how diseases spread in a population.
We examine how
the network structure affects the propagation of a recession
throughout the world. For example, if there is a one percentage
decrease of the GDP of the USA,
by how much does the total GDP of world excluding the USA decrease due to the spread of recession from the USA?
We investigate the five most recent global recessions including the 2007-2009 crisis. We calculate the ratio of GDP change (percentage) of world excluding the USA to the GDP change (percentage) of USA in each recession as a function of
 the CCC value in each recession, see Fig.\ \ref{sim}(a). We observe
 that in more recent recessions with less hierarchical structure of trade
 network, a recession in the USA has a stronger impact on the rest
of the world. This result indicates a strong positive correlation between lack of hierarchical structure and severity of recession impact.

We also perform an impulse response analysis of the
vector autoregression (VAR) model to analyze the time evolution of
recession \cite{carvalho2009,koop1996,abey}. We 
explore the possible underlining causal links between
lack of hierarchical structure and severity of recession.
A recession is assumed to start in the USA.
The US GDP is initially
reduced by
the maximal GDP decline during the recession, e.g.\ the maximal
 quarterly US GDP decline was 5.4\%
S in the 2008--2009 recession \cite{economist}. 
The export from country $i$ to country $j$, $ X_{ij}$,
is updated by a factor of ratio of GDP of country $j$ at time $t$,
$Y_j(t)$, and $t-1$, $Y_j(t-1)$) \cite{abey}. Thus,
$X_{ij}(t) = X_{ij}(t-1) Y_j (t)/ Y_j(t-1)$. Then the GDP
of country $i$ is updated by $Y_i (t+1)=Y_i
(t)+P_i(X_i(t)/X_i (t-1)-1)$. Where $P_i=X_i/Y_i$ is the
ratio of export to GDP for country $i$.
 The GDP of each country decreases until steady
state is reached, at which point the simulation is terminated.
 We calculate
the world GDP change as $(\sum_i Y_i^{\rm steady}- \sum_i
Y_i^{\rm initial})/\sum_i Y_i^{\rm initial}$.
We observe how the crisis spreads globally and measured the GDP loss during crisis.


The impulse response analysis results support that the severity of the 2008-2009 recession may be due to
loss of hierarchical structure in the global trade network.  Lack of hierarchical structure makes the world trade network less resistant to recession, as
observed from Fig.\ \ref{sim}(b).
We believe this increased sensitivity is due to a loss
of modular or hierarchical structure in the world trade network, see Fig.\ \ref{ccc1}. As
an example, the impact of a recession on the GDP is more severe in
2006 than in 1968, by a factor of 5.7.  Interestingly, after this
calculation was carried out, an estimate of the ratio
of the reduction of GDP in
2009 to the average reduction over past recessions equaling 6
was reported \cite{economist}.

Evolutionary theory has shown that systems under environmental
perturbation not only increase their modularity, but also increase
their response function to perturbations \cite{jun,he2009}.
In the present context, this would imply that as trade has been globalized,
and the CCC reduced, the rate of recovery from recession should decrease.
We consider this phenomenon in the world
trade network, using the VAR model.
After the
system reaches steady state following the reduction to the USA GDP, we impose a
positive impulse to restore the USA GDP to its initial value.
The world GDP recovers, at a rate that depends on the hierarchical structure of the
trade network.
We observe that when the trade network has greater hierarchical structure,
 indicated by a
larger CCC value, the trade network recovers more quickly from
recession, as shown in the insert figure of Fig.\ \ref{sim}(b).


We have used the concept of viewing the world trade network as
defining a geometry in trade space and the idea of projecting this
geometry to the best tree-like topology to define the hierarchy
in the world trade network.  With that necessary mathematical
prolegomena,  we introduced the world trade network
as an evolving system.  Physics of evolution in changing environments
was then used to predict that
the world trade network is more sensitive and recovers
more slowly from evolutionary shocks now than it did 40 years ago,
because globalization has reduced hierarchical structure in the
world trade network.  We also predict that
recession-induced change to the world trade network should lead to
a temporarily \emph{increased} hierarchical structure of the global trade
network.   These predictions, contrary to standard economic
thinking, were born out by our study of the
world trade data since 1969.


\bibliography{jiankui}

\clearpage

\end{document}